\documentclass[aps,prl,twocolumn,amsfonts,showpacs,superscriptaddress,longbibliography]{revtex4-2}
\usepackage{verbatim,epsfig,amsmath,amssymb,bm,epsf,graphicx,psfrag,bbold,amsthm,amsfonts,dsfont}
\usepackage[bottom]{footmisc}
\usepackage{hyperref}
\usepackage{cleveref} 
\usepackage{enumitem}
\usepackage{framed}
\usepackage{mathrsfs}
\usepackage{esint}
\setlist{nosep}
\usepackage{placeins}
\usepackage[all]{xy}
\usepackage{color}
\usepackage[utf8]{inputenc}
\usepackage{float}
\usepackage{natbib}
\usepackage{tikz-cd}
\usepackage{verbatim}
\usepackage{leftidx}
\usepackage[normalem]{ulem}

\usepackage{notes2bib}

\newcommand{\ket}[1]{\mbox{$ | #1 \rangle $}}

\newcommand{\cT}{\mathcal{T}}

\newcommand{\cK}{\mathcal{K}}
\newcommand{\cN}{\mathcal{N}}

\newcommand\bZ {{\mathbb Z}}

\newcommand\beq {\begin{equation}}
\newcommand\eeq {\end{equation}}
\newcommand\beqa {\begin{equatiobn}\begin{array}}
\newcommand\eeqa {\end{array}\end{equation}}
\newcommand\bal {\begin{align}}
\newcommand\eal {\end{align}}
\newcommand{\bea}{\begin{eqnarray}}
\newcommand{\eea}{\end{eqnarray}}
\newcommand{\ii}{\text{i}}

\newcommand{\outerproduct}[2]{|#1\rangle\langle #2|}

\newcommand{\ztwo}{\mathbb{Z}_2}

\theoremstyle{plain}

\theoremstyle{definition}

\theoremstyle{remark}

\begin{document}

\title{Boundary supersymmetry of 1+1 d fermionic SPT phases}

\author{Abhishodh Prakash}
\email{abhishodh.prakash@icts.res.in (he/him/his)}
\affiliation{International Centre for Theoretical Sciences (ICTS-TIFR),
Tata Institute of Fundamental Research,
Shivakote, Hesaraghatta Hobli,
Bengaluru 560089, India}
\author{Juven Wang}
\email{jw@cmsa.fas.harvard.edu (he/him/his)}
\affiliation{Center of Mathematical Sciences and Applications, Harvard University,  Cambridge, MA 02138, USA}

\begin{abstract}
	We prove that the boundaries of all non-trivial 1+1 dimensional intrinsically fermionic symmetry-protected-topological phases, protected by finite on-site symmetries (unitary or anti-unitary), are supersymmetric quantum mechanical systems. This supersymmetry does not require any fine-tuning of the underlying Hamiltonian, arises entirely as a consequence of the boundary 't Hooft anomaly that classifies the phase and is related to a `Bose-Fermi' degeneracy different in nature from other well known degeneracies such as Kramers doublets.
\end{abstract}

\maketitle 

\noindent Topological phases of quantum matter are exotic insulating phases which fall outside the paradigm of spontaneous symmetry breaking. The simplest class of these phases are symmetry-protected-topological (SPT) phases which are almost indistinguishable from the \emph{trivial} phase in the bulk but have marvellous properties on the boundaries~\cite{HasanKane_TI_RevModPhys.82.3045,QiZhang_RevModPhys.83.1057,SatoTSC__2017} such as symmetry protected gaplessness, spectral degeneracies and persistent ordering~\cite{Komargodski_2019}. Well-known examples of SPT phases are topological insulators~\cite{HasanKane_TI_RevModPhys.82.3045,QiZhang_RevModPhys.83.1057}, topological superconductors~\cite{SatoTSC__2017,QiZhang_RevModPhys.83.1057} and the Haldane spin chain~\cite{AKLT_PhysRevLett.59.799}. SPT phases are a subset of the more general \emph{invertible} phases whose anomalous boundaries may survive even if symmetries are explicitly broken. For a particular dimension of space and global symmetries, invertible phases of bosons or fermions form an \emph{Abelian group} of which SPT phases are a subgroup. The program of \emph{classifying} invertible and SPT phases~\cite{Xiong_2018,GaiottoJohnson-Freyd_2019} has uncovered fascinating algebraic structures present in the space of quantum many-body systems through a vibrant collaboration between areas of theoretical physics and mathematics. The non-trivial nature of the boundaries of SPT phases arises from an \emph{anomalous}~\cite{Wen_anomalies_PhysRevD.88.045013} realization of global symmetries on the boundary degrees of freedom~\cite{ElseNayak_PhysRevB.90.235137}. This \emph{'t Hooft anomaly} is manifested in the form of emergent, extended symmetries~\cite{WangWenWitten_PhysRevX.8.031048} and fractional quantum numbers~\cite{VishwanathSenthil_3dBTI_PhysRevX.3.011016}. In one spatial dimension, the anomalous boundary manifests itself by symmetries being represented \emph{projectively}~\cite{ChenGuWen_1d_PhysRevB.84.235128,FidkowskiKitaev_Z8_PhysRevB.83.075103,FidkowskiKitaev_interactions_PhysRevB.81.134509}. In this Letter, we present a universal property of non-trivial 1+1 d \emph{intrinsically fermionic} SPT phases i.e. which cannot be interpreted as bosonic SPT phases stacked with gapped fermions protected by finite on-site symmetries. We prove that as a consequence of the boundary anomaly, i.e. projectively represented fermionic symmetries, any symmetry preserving boundary Hamiltonian of any member of any such phase is a supersymmetric quantum mechanical system~\cite{CooperKhareSukhatme_SUSYQM1995,WITTEN_SUSYQM_1981513,WITTEN_SUSYQM_1982253}. 

Originally introduced as a resolution to certain issues in fundamental and particle physics~\cite{Witten_IntroSUSY_1983}, supersymmetry (SUSY) has served as a powerful theoretical tool to uncover aspects of quantum field theories~\cite{Seiberg_1995} as well as a way to elegantly prove mathematical theorems using techniques from physics~\cite{Witten_SUSY_Morse_:1982im}. Although signatures of space-time SUSY in particle interactions have not yet been detected in particle colliders~\cite{Autermann_2016}, proposals exist for their emergence and detection in condensed matter~\cite{Lee2010tasi,GroverVishwanath_SUSY_280,Hsieh_PhysRevLett.117.166802,Qi_SUSYPhysRevLett.102.187001,Yao_PhysRevLett.114.237001,YaoLieaau1463,Yao_PhysRevLett.118.166802,Yao_PhysRevLett.119.107202} and cold-atomic systems~\cite{Tomka_SUSYColdatom_2015NatSR...513097T,Lahrz_SUSYColdatom_PhysRevA.96.043624,YuKun_SUSY_PhysRevLett.100.090404,BradlynGromov_SUSY_PhysRevA.93.033642}. To the best of our knowledge, all these proposals require some combination of a) fine tuning of parameters, b) exotic bulk symmetries or c) special local Hilbert space structure. 

Our work provides a generic and robust setting for the emergence and detection of the simplest version of SUSY- supersymmetric quantum mechanics (SUSY QM)~\cite{CooperKhareSukhatme_SUSYQM1995,WITTEN_SUSYQM_1981513,WITTEN_SUSYQM_1982253} without any fine-tuning, in one-dimensional fermion systems with symmetries and Hilbert spaces accessible to current experimental efforts~\cite{HasanKane_TI_RevModPhys.82.3045,QiZhang_RevModPhys.83.1057} such as spinful fermions with time-reversal symmetry. This is because the emergent SUSY in our case is a consequence of the boundary 't Hooft anomaly- a robust property of entire phases of matter and consequently can thus be observed in any member of any of these phases. 

In the first part of this Letter, we demonstrate the emergence of boundary SUSY using the example of 1+1 d time-reversal invariant topological superconductors~\cite{FidkowskiKitaev_Z8_PhysRevB.83.075103,FidkowskiKitaev_interactions_PhysRevB.81.134509}. In the second, we place the example in a general setting by proving a theorem which establishes that boundary SUSY is a generic feature of all non-trivial 1+1 d intrinsically fermionic SPT phases with finite on-site symmetries.

\begin{figure}[!htbp]
	\centering	
	\includegraphics[width=0.43\textwidth]{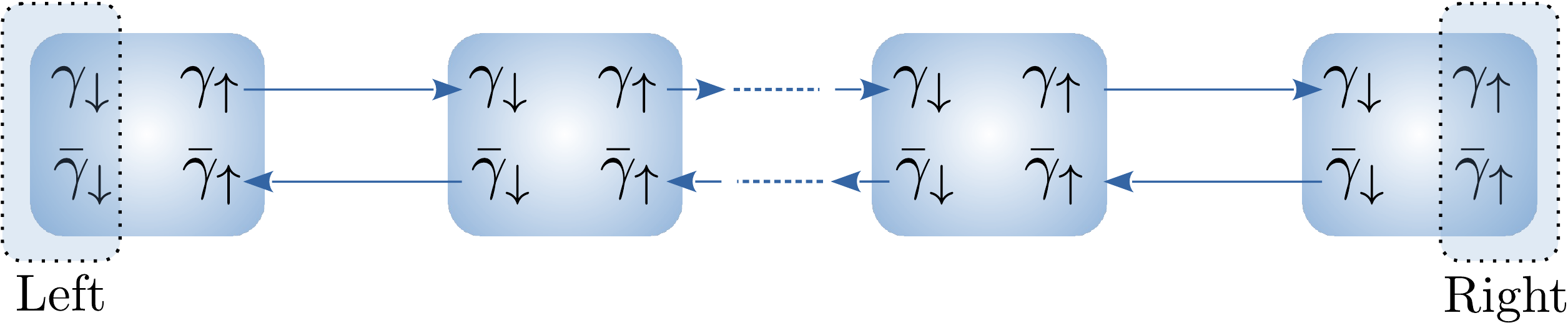}
	\caption{Schematic representation of the Hamiltonian in~\cref{eq:BDI} with $\ztwo^T \times \ztwo^f$ symmetry.\label{fig:BDI}}
\end{figure}

\medskip
\noindent \emph{\textbf{Boundary SUSY of topological superconductors}}:
 In refs~\cite{FidkowskiKitaev_Z8_PhysRevB.83.075103,FidkowskiKitaev_interactions_PhysRevB.81.134509}, Fidkowski and Kitaev established that invertible phases of 1+1 d interacting topological superconductors with time-reversal symmetry $\cT$ satisfying $\cT^2 = +\mathds{1}$ form a $\bZ_8$ group. Of these, the phases labeled by even numbers, $\nu = 2,4,6$, forming a $\bZ_4$ subgroup, are SPT phases i.e. require symmetry protection and are trivial in its absence. Let us consider the  phase corresponding to $\nu = 2$. A representative model in this phase is two copies of Kitaev's Majorana chain~\cite{Kitaev_majoranachain_2001,Gu_Neutrinos_PhysRevResearch.2.033290}. Using a local change of basis~\cite{APWangWei_Unwinding_PhysRevB.98.125108}, for a system of size $L$, the Hamiltonian and symmetry operators (time-reversal, $\cT$ and fermion parity, $P_f$) can be written as 
\begin{align}
H &= -\ii \sum_{j=1}^L \left(\gamma_{\uparrow,j} \gamma_{\downarrow,j+1}- \bar{\gamma}_{\uparrow,j} \bar{\gamma}_{\downarrow,j+1}\right), \label{eq:BDI} \\
\cT & = 
\mathcal{K} ~\prod_{j=1}^L 
(\gamma_{\downarrow,j} \bar{\gamma}_{\uparrow,j}),~P_f = \prod_{j=1}^L \left(\ii \bar{\gamma}_{\downarrow,j} \gamma_{\downarrow,j}\right) \left(\ii \bar{\gamma}_{\uparrow,j} \gamma_{\uparrow,j}\right) \nonumber.
\end{align}
The Hilbert space on each site consists of two complex fermions (labeled $\sigma=\uparrow, \downarrow$) and represented using four Majorana operators $\left(\gamma_{\sigma}, \bar{\gamma}_{\sigma}\right)$. We work with a basis such that $\gamma_{\sigma}$ are real and symmetric, and $\bar{\gamma}_{\sigma}$ are imaginary and antisymmetric~\cite{Clifford}. Under symmetries, the Majorana operators transform as
\begin{align}
\cT &: \gamma_{\sigma} \mapsto \tau^z_{\sigma \sigma'} \gamma_{\sigma'},~~ \bar{\gamma}_{\sigma} \mapsto \tau^z_{\sigma \sigma'} \bar{\gamma}_{\sigma'},~~ \ii \mapsto -\ii \label{eq:Bulk T} \\
P_f &: \gamma_{\sigma} \mapsto ~-\gamma_{\sigma},~~~~ \bar{\gamma}_{\sigma} \mapsto ~- \bar{\gamma}_{\sigma},~~~\ii \mapsto~~ \ii \label{eq:Bulk Pf},
\end{align}
where, $\tau^z$ is the Pauli-Z matrix. The symmetry operators satisfy
\begin{equation}
\cT^2 = P_f^2 =\mathds{1},~~\cT P_f = P_f \cT. \label{eq:Z2 Z2 bulk}
\end{equation} 
and generate a $\ztwo^T \times \ztwo^f$ group. With open boundary conditions, we get a pair of unpaired Majorana modes on each end as shown in \cref{fig:BDI} which results in a four-fold ground state degeneracy. We can now consider the time-reversal and fermion parity operators restricted to the Hilbert space of one of the boundaries (say left):
\begin{align}
\hat{\mathcal{T}} = \cK \gamma,~~\hat{P}_{f} = i \bar{\gamma} \gamma.
\end{align}
We have relabeled the boundary modes $\gamma_{\downarrow,1} \equiv \gamma$, $\bar{\gamma}_{\downarrow,1} \equiv \bar{\gamma}$ for convenience. Throughout this Letter, we use operators with a hat to label boundary operators (symmetry and Hamiltonian) to distinguish them from those in the bulk.
It can be checked that the boundary symmetry generators satisfy
\begin{equation}
\hat{\cT}^2 = \hat{P}_f^2 =\mathds{1},~~\hat{\cT} \hat{P}_f = - \hat{P}_f \hat{\cT}, \label{eq:D8 extension}
\end{equation} 
and thus form a non-trivial \emph{projective representation}~\footnote{Briefly, a projective representation of a group $G$ is an assignment of a matrix $V(g)$ for each element $g_i \in G$ that satisfies group composition \emph{upto an overall complex phase} $V(g_1) V(g_2) = e^{i \omega(g_1,g_2)} V(g_1.g_2)$. The projective representation is non-trivial if the complex phases $e^{i \omega(g_1,g_2)}$ cannot be removed by re-phasing $V(g) \mapsto e^{i \theta(g)} V(g)$. The distinct possible projective representations form an Abelian group $H^2(G,U(1)_T)$.} of the bulk $\ztwo^T \times \ztwo^f$ group~\cref{eq:Z2 Z2 bulk}. The action of symmetries on the boundary operators is
\begin{align}
\hat{\cT} &: \gamma \mapsto +\gamma,~~ \bar{\gamma} \mapsto +\bar{\gamma},~~ \ii \mapsto -\ii \\
\hat{P}_f &: \gamma \mapsto -\gamma,~~ \bar{\gamma} \mapsto - \bar{\gamma},~~\ii \mapsto~~ \ii ,
\end{align}
and the only possible boundary Hamiltonian consistent with symmetries is 
\begin{equation}
\hat{H} = c\mathds{1}, \label{eq:trivial boundary BDI}
\end{equation}
where, $c$ is a real number which we choose to be positive without any loss of generality. Observe that the spectrum of~\cref{eq:trivial boundary BDI} has a Bose-Fermi degeneracy consisting of one bosonic ($\hat{P}_f=+1$) and one fermionic ($\hat{P}_f=-1$) eigenstate with the same energy ($E=c$). This is a consequence of the fact that the boundary time-reversal operator \emph{anti-commutes} with fermion parity. We can define two Hermitian \emph{supercharges}, $\hat{Q}_+ \equiv \sqrt{c} \gamma$ and $\hat{Q}_- \equiv  \sqrt{c} \bar{\gamma}$ which satisfy the $\cN=2$ SUSY QM algebra,
\begin{equation}
\{\hat{Q}_\alpha,\hat{Q}_\beta\} = 2\hat{H} \delta_{\alpha \beta},~[\hat{H},\hat{Q}_\alpha]= \{\hat{P}_f,\hat{Q}_\alpha\}=0. \label{eq:N = 2 SUSY QM}
\end{equation}
The action of time-reversal on the supercharges can also be determined and is best seen on the \emph{complex} supercharges, $Q \equiv \hat{Q}_+ + \ii \hat{Q}_-$ and $\bar{Q}\equiv \hat{Q}_+ - \ii \hat{Q}_-$,
\begin{equation}
\hat{\cT}: \begin{pmatrix}
Q \\
\bar{Q}
\end{pmatrix} \mapsto \begin{pmatrix}
\bar{Q} \\
Q
\end{pmatrix}. \label{eq:R symmery}
\end{equation}
Remarkably, this emergent SUSY is a feature of \emph{any} member of the $\nu=2$ phase, not just the model of~\cref{eq:BDI}. To see this, let us consider a model in the same phase with a more complex boundary Hilbert space.
\begin{figure}[!htbp]
	\centering	
	\includegraphics[width=0.43\textwidth]{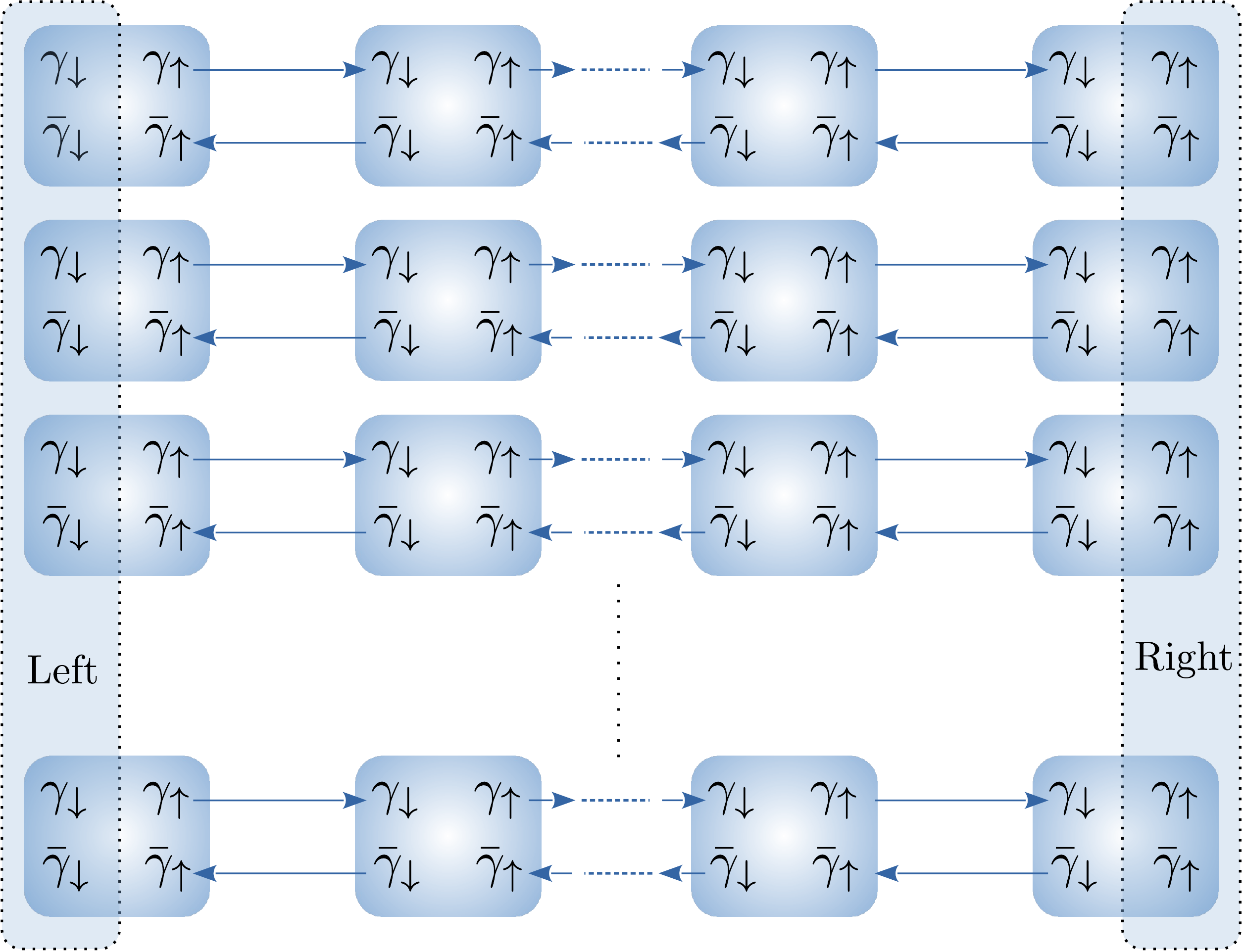}
	\caption{Schematic representation of the Hamiltonian in~\cref{eq:BDI_layer} with $\ztwo^T \times \ztwo^f$ symmetry.\label{fig:Layered_BDI}}
\end{figure}
The $\bZ_8$ classification~\cite{FidkowskiKitaev_Z8_PhysRevB.83.075103,FidkowskiKitaev_interactions_PhysRevB.81.134509} tells us that we can stack an additional $4N$ copies of the Hamiltonian of \cref{eq:BDI} onto the original one without changing the phase of matter. Doing this gives us the following Hamiltonian and symmetry operators:
\begin{align}
H &= -\ii \sum_{j=1}^L \sum_{a=1}^{4N + 1} \left(\gamma_{\uparrow,a,j} \gamma_{\downarrow,a,j+1}- \bar{\gamma}_{\uparrow,a,j} \bar{\gamma}_{\downarrow,a,j+1}\right), \label{eq:BDI_layer} \\
\cT  &= \cK\prod_{a=1}^{4N+1}\prod_{j=1}^L
(\gamma_{\downarrow,a,j} \bar{\gamma}_{\uparrow,a,j}),  \\
P_f &=\prod_{a=1}^{4N+1} \prod_{j=1}^L
 \left(  \ii \bar{\gamma}_{\uparrow,a, j} \gamma_{\uparrow,a, j} \right) \left(  \ii \bar{\gamma}_{\downarrow,a, j} \gamma_{\downarrow,a, j} \right). 
\end{align}
The action of symmetry on the Majorana operators is the same as shown in \cref{eq:Bulk T,eq:Bulk Pf}. With open boundary conditions, as shown in \cref{fig:Layered_BDI}, the Hilbert space on the left end now consists of $8N+2$ unpaired Majorana operators $\left(\gamma_{1},\ldots ,\gamma_{4N+1},\bar{\gamma}_{1}, \ldots ,\bar{\gamma}_{4N+1}\right)$. We have again relabeled the boundary modes $\gamma_{\downarrow,a,1} \equiv \gamma_a$, $\bar{\gamma}_{\downarrow,a,1} \equiv \bar{\gamma}_a$ for convenience. The boundary symmetry generators  as well as their action on boundary Majorana operators are:
\begin{align}
\hat{\cT} &= \cK \prod_{a=1}^{4N+1} \gamma_a , ~~~ \hat{P}_f= \prod_{a=1}^{4N+1} \ii \bar{\gamma}_a \gamma_a,\\
\hat{\cT} &: \gamma_a \mapsto +\gamma_a,~~ \bar{\gamma}_a \mapsto +\bar{\gamma}_a,~~ \ii \mapsto -\ii, \\
\hat{P}_f &: \gamma_a \mapsto -\gamma_a,~~ \bar{\gamma}_a \mapsto - \bar{\gamma}_a,~~\ii \mapsto~~ \ii .
\end{align}
It can be checked that $\hat{\cT}$ and $\hat{P}_f$ form the same projective representation as~\cref{eq:D8 extension}. However, the symmetry-allowed boundary Hamiltonian can be more complex and can include any operator consisting of Majoranas coupled in multiples of four (quartic, octonic, $\ldots$):
\begin{multline}
\hat{H} = \sum_{a,b,c,d}J^1_{abcd}~ \gamma_a \gamma_b \gamma_c \gamma_d + J^2_{abcd}~ \bar{\gamma}_a \gamma_b \gamma_c \gamma_d + \ldots\\
\sum_{a,b,c,d,e,f,g,h} K^1_{abcdefgh}~ \gamma_a \gamma_b \gamma_c \gamma_d \gamma_e \gamma_f \gamma_g \gamma_h + \ldots \label{eq:SYK}
\end{multline}
With only quartic couplings and $J^\alpha_{abcd}$ chosen from a random distribution,  this is the Sachdev-Ye-Kitaev  (SYK) model~\cite{YouLudwigXu_SYKSPT_PhysRevB.95.115150,Kitaev_SYK,Rosenhaus_SYK2019}. The projective boundary symmetry representation~\cref{eq:D8 extension} again enforces the spectrum to have Bose-Fermi degeneracy. To see this, observe that if $\ket{\mu,+}$ is some bosonic eigenstate with a non-zero eigenvalue $E_\mu$, \cref{eq:D8 extension} ensures that $\ket{\mu,-} = \cT \ket{\mu,+}$ is a fermionic eigenstate with the same eigenvalue. This means, in the absence of any other accidental degeneracies (ensured by $J^\alpha_{abcd}$ etc being sufficiently random), the Hamiltonian of \cref{eq:SYK} can be written in diagonal form as
\begin{align}
\hat{H} &= \sum_{\mu} E_\mu \left(\outerproduct{\mu,+}{\mu,+} + \outerproduct{\mu,-}{\mu,-}\right),\\
\hat{H} &\ket{\mu,\pm} = E_\mu \ket{\mu,\pm}, ~~\hat{P}_f \ket{\mu,\pm} = \pm \ket{\mu,\pm}. \nonumber
\end{align}
If we shift energies so as to ensure all $E_\mu$ are positive, as shown by Behrends and B\'{e}ri~\cite{BehrendsBeri_SUSYSYK_PhysRevLett.124.236804}, we can again define two supercharges
\begin{align}
\hat{Q}_{+} &= \sum_{\mu} \hat{Q}_\mu + \hat{Q}^\dagger_\mu, ~~~~~\hat{Q}_{-} =\sum_{\mu} \ii \left( \hat{Q}_\mu - \hat{Q}^\dagger_\mu \right), \\
&\text{where, } \hat{Q}_\mu = \sqrt{E_\mu } \outerproduct{\mu,+}{\mu,-}, \nonumber
\end{align}
which generate the same $\cN = 2$ SUSY QM algebra as~\cref{eq:N = 2 SUSY QM} as well as the action of time-reversal symmetry on the supercharges shown in~\cref{eq:R symmery}. A similar story also exists for the $\nu=6$ phase generated by $4N+3$ copies of Hamiltonian~\cref{eq:BDI} whose boundary also corresponds to a $\cN=2$ SUSY QM system but with a different action of symmetries.

The fact that the SYK model can serve as a boundary Hamiltonian for topological superconductors was pointed out by You, Ludwig and Xu~\cite{YouLudwigXu_SYKSPT_PhysRevB.95.115150} and the SUSY nature of the `standard' SYK model was recently studied by Behrends and B\'{e}ri~\cite{BehrendsBeri_SUSYSYK_PhysRevLett.124.236804}. The main result of this work is a proof that this is true of \emph{all} systems belonging to \emph{all} non-trivial 1+1 d intrinsically fermionic SPT phases protected by \emph{any} finite on-site symmetry, irrespective of the choice of boundary Hamiltonians.

\medskip 
\noindent \emph{\textbf{Proof of the general theorem:}}  We now prove the main theorem of this Letter:
\medskip

\noindent \emph{Any boundary Hamiltonian of a system belonging to a non-trivial 1+1 d SPT phase protected by finite on-site unitary or anti-unitary symmetries can be expressed as a supersymmetric quantum mechanical system if and only if the SPT phase is intrinsically fermionic. }
\medskip

 Let us first carefully state the setting we are considering and clarify some terminology. By non-trivial 1+1 d SPT phases, we mean classes of Hamiltonians in one spatial dimension which have a unique ground state with closed boundary conditions but have ground state degeneracies in the presence of any open boundary conditions preserving symmetry. If the global symmetries are explicitly broken, the boundary can be gapped and the phase becomes trivial. We are interested in fermionic systems whose \emph{total symmetry group} $G_f$ consists of fermion parity $P_f \equiv (-\mathds{1})^{N_f}$ which commutes with all other symmetry operators. In other words, $\ztwo^f\cong \{\mathds{1},P_f\}$ is in the \emph{center} of the group $G_f$~\cite{Hamermesh:100343}. We also define the \emph{bosonic symmetry group} as the \emph{quotient} $G_b \cong G_f/\ztwo^f$~\footnote{$G_b$ could be a subgroup of $G_f$ or not. e.g.: $\cT^2=\mathds{1}$ time-reversal symmetry has $G_f\cong  \ztwo^f \times \ztwo^T$ and $\cT^2 = P_f$ time-reversal symmetry has $G_f \cong \bZ_4^{f,T}$. Both the groups have $G_b \cong \ztwo^T$ but the former contains $G_b$ as a subgroup while the latter does not.}. We assume that $G_f$ forms a finite group and all operators in $G_f$, which can be unitary or anti-unitary, are on-site~\cite{ChenGuWen_1d_PhysRevB.84.235128} i.e. can be written as a product of operators which act on a finite number of local degrees of freedom, possibly followed by complex conjugation. The boundary symmetry operators of such an SPT phase, $\hat{G}_f$ forms a \emph{projective representation} of $G_f$ and is classified by the \emph{second group cohomology group} $H^2(G_f,U(1)_T)$~\cite{KapustinTurzilloYou_SpinTFT_MPS_PhysRevB.98.125101,KapustinThorngren_2017,TurzilloYou_FermionicMPS_PhysRevB.99.035103,FidkowskiKitaev_Z8_PhysRevB.83.075103} which in turn also classifies the 1+1 d fermionic SPT phase~\footnote{Invertible phases that are non-trivial even in the absence of symmetries are not included in the $H^2(G_f,U(1)_T)$ classification. For example, the the $\nu=1$ member of time-reversal invariant topological superconductors with $\cT^2=\mathds{1}$ of which Kitaev's Majorana chain is a member~\cite{Kitaev_majoranachain_2001}. See \cite{KapustinTurzilloYou_SpinTFT_MPS_PhysRevB.98.125101,TurzilloYou_FermionicMPS_PhysRevB.99.035103,KapustinThorngren_2017} for more details on the classification of such phases.}. An SPT phase is \emph{intrinsically} fermionic unless the classification reduces to $H^2(G_b,U(1)_T)$~\cite{ChenGuWen_1d_PhysRevB.84.235128} in which case the phase can be thought of as a non-trivial bosonic SPT phase stacked with trivial gapped fermions. 

To begin, let us consider a Hamiltonian $H$ belonging to a non-trivial SPT phase protected by total symmetry $G_f$. This means that the boundary symmetry operators $\hat{G}_f$ are projectively realized which forbids any boundary Hamiltonian $\hat{H}$ invariant under $\hat{G}_f$ from acquiring a unique ground state invariant. We prove the main theorem in two steps:

\medskip

\noindent \textbf{(I)} We prove that an SPT phase is not intrinsically fermionic if and only if $\hat{P}_f$  commutes with all elements of $\hat{G}_f$.\\
\noindent \textbf{(II)} We prove that if $\hat{P}_f$ does not commute with all elements of $\hat{G}_f$, then $\hat{H}$ is supersymmetric.

\medskip

\noindent \textbf{Proof of (I)}: We first prove this using the formal results of classification of fermionic SPT phases~\cite{GuWen_fSPT_PhysRevB.90.115141,WangGu_fSPT_PhysRevX.8.011055,WangGu_PhysRevX.10.031055,KapustinTurzilloYou_SpinTFT_MPS_PhysRevB.98.125101,TurzilloYou_FermionicMPS_PhysRevB.99.035103,KapustinThorngren_2017} and then provide a physical interpretation.  

The classification of fermionic SPT phases, given by $H^2(G_f,U(1)_T)$ can be specified by two pieces of data- $\alpha$ and $\beta$ (see refs~\cite{TurzilloYou_FermionicMPS_PhysRevB.99.035103,KapustinTurzilloYou_SpinTFT_MPS_PhysRevB.98.125101,KapustinThorngren_2017} for details). If and only if $\beta \in  H^1(G_b,\ztwo)$ is trivial, the classification reduces to $\alpha \in H^2(G_b,U(1)_T)$ and the phase is not intrinsically fermionic. Now, $\beta$, which assigns a $\ztwo$ element (either 0 or 1) to each element of $G_b$ also encodes the action of fermion parity on each operator $\hat{V}_g \in \hat{G}_f$~\cite{TurzilloYou_FermionicMPS_PhysRevB.99.035103}. 
\begin{equation}
\hat{P}_f \hat{V}_g \hat{P}_f^{-1} = (-1)^{\beta\left(\pi(g)\right)}~ \hat{V}_g. \label{eq:beta}
\end{equation}
Here, $\pi$ is the surjective map from the elements of the group $g \in G_f$ to the quotient $G_b \cong G_f/\ztwo^f$. Now, if $\beta$ is trivial i.e. assigns $0$ to every element of $G_b$, we see from \cref{eq:beta} that this precisely means that fermion parity commutes with all elements of $\hat{G}_f$. Furthermore, the argument also works in reverse- if $\hat{P}_f$ commutes with all the elements of $\hat{G}_f$, \cref{eq:beta} tells us that $\beta$ must be trivial and the SPT phase is not intrinsically fermionic.

Let us now understand what this means physically. If the boundary fermion parity $\hat{P}_f$ commutes with all other elements of $\hat{G}_f$, we can simply add $\hat{P}_f$ to the boundary Hamiltonian, $\hat{H}$ without breaking any symmetries to get
\begin{equation}
\hat{H}(\lambda) = \hat{H} -\lambda \hat{P}_f.
\end{equation}
Now, we can take $\lambda$ to be large and positive so as to ensure that the low energy states are all bosonic i.e. have $P_f=+1$. The effective representation of symmetry generators on this low energy bosonic subspace is $\hat{G}_b  \cong \hat{G}_f/\ztwo^f$. Since the SPT phase is assumed to be non-trivial, the projection of $\hat{H}(\lambda)$ to the $P_f=+1$ sector cannot have a unique ground state. This means $\hat{G}_b$ is a projective representation classified by some non-trivial element of $H^2(G_b,U(1)_T)$. Thus, the SPT phase is not intrinsically fermionic. 

\medskip
\noindent \textbf{Proof of (II)}: Recall that the fermion parity $\hat{P}_f$ grades operators as bosonic or fermionic 
\begin{equation}
\hat{P}_f \hat{O}_b \hat{P}_f^{-1} = + \hat{O}_b, ~~~\hat{P}_f \hat{O}_f \hat{P}_f^{-1} = - \hat{O}_f.
\end{equation}
We will assume that the symmetry operators of $\hat{G}_f$ all have \emph{definite fermion parity} i.e. are either bosonic or fermionic. This is indeed the case for projective representations of fermionic symmetries as shown in \cref{eq:beta}~\cite{TurzilloYou_FermionicMPS_PhysRevB.99.035103,KapustinTurzilloYou_SpinTFT_MPS_PhysRevB.98.125101,KapustinThorngren_2017}. If $\hat{P}_f$ does not commute with all symmetry operators and there exist some symmetry operators $\hat{V}_g$ that anti-commutes with $\hat{P}_f$, the eigenstates of any Hamiltonian $\hat{H}$ invariant under $\hat{V}_g$ with non-zero eigenvalues has Bose-Fermi degeneracy similar to \cref{eq:trivial boundary BDI,eq:SYK}. To see this, observe that if $\ket{\mu}$ is a bosonic eigenstate with non-zero eigenvalue $E_\mu$, $\hat{V}_g\ket{\mu}$ is a fermionic eigenstate with the same eigenvalue. In diagonal form, $\hat{H}$ can be written as
\begin{align}
\hat{H} &= \sum_{\mu,a_\mu} E_\mu \left(\outerproduct{\mu,a_\mu,+}{\mu,a_\mu,+} + \outerproduct{\mu,a_\mu,-}{\mu,a_\mu,-}\right), \nonumber \\
\hat{H} &\ket{\mu,a_\mu,\pm} = E_\mu \ket{\mu,a_\mu,\pm},~~ \hat{P}_f \ket{\mu,a_\mu,\pm} = \pm \ket{\mu,a_\mu,\pm}. \nonumber
\end{align}
The label $a_\mu$ keeps track of additional degeneracies of $E_\mu$ arising from the specific nature of the symmetries (e.g.: Kramers degeneracy, non-Abelian symmetries etc.). After shifting energies to make all $E_\mu$ positive, we can always define \emph{at least} two supercharges
\begin{align}
\hat{Q}_{+} &= \sum_{\mu} \hat{Q}_\mu + \hat{Q}^\dagger_\mu, ~~~~~\hat{Q}_{-} =\sum_{\mu} \ii\left( \hat{Q}_\mu - \hat{Q}^\dagger_\mu \right), \\
&\text{where, } \hat{Q}_\mu = \sum_{a_\mu} \sqrt{E_\mu } \outerproduct{\mu,a_\mu,+}{\mu,a_\mu,-}, \nonumber
\end{align}
which satisfy the $\cN=2$ SUSY algebra of \cref{eq:N = 2 SUSY QM}. We stress that we cannot rule out the possibility of a larger number of supercharges.

\medskip

\noindent Contraposing (\textbf{I}), we can establish that an SPT phase is intrinsically fermionic if and only if the boundary fermion parity $\hat{P}_f$ does not commute with all boundary symmetry operators. Combined with \textbf{(II)}, this implies that the boundary Hamiltonian, $\hat{H}$ is a supersymmetric quantum mechanical system. This completes the proof.

\medskip

\noindent \textbf{\textit{Comments}}: We briefly comment on how our results relate to existing ones. First, we note that the manifestation of the boundary anomaly of SPT phases in the form of spectral degeneracies is well-known. While for bosons, the boundary degeneracy corresponds to fractional quantum numbers~\cite{AKLT_PhysRevLett.59.799} or Kramers doublets~\cite{PollmannTurnerBergOshikawa_PhysRevB.81.064439}, for intrinsically fermionic systems, the degeneracy is different and of the Bose-Fermi kind discussed above. When the system possesses time-reversal symmetry, this degeneracy is sometimes referred to as fermionic Kramers doublets~\cite{Metlitski2014interaction} and in Ref~\cite{Qi_SUSYPhysRevLett.102.187001}, this degeneracy is itself referred to as supersymmetry. 

The main result of our Letter is the proof of the existence of supercharges which generate a SUSY algebra. This has not been identified before to the best of our knowledge. In addition to producing the Bose-Fermi spectral degeneracy, SUSY can constrain the dynamics of the system and imprint signatures in static and dynamical correlation functions~\cite{FuGaiottoMaldacenaSachdevo_SUSY_SYKPhysRevD.95.026009,BehrendsBeri_SUSYSYK_PhysRevLett.124.236804}. We leave the study of these consequences for future work.

\medskip
\noindent \emph{\textbf{Summary and outlook:}} We have described a general setting for the emergence of supersymmetric quantum mechanics. Since this applies to a large class of physical systems belonging to several phases with no fine tuning, we expect it is a promising avenue for experimental detection of SUSY in cold-atom as well as condensed matter systems. Our work also opens up several questions for theoretical investigation. First, we expect that some version of emergent SUSY should also be present on the boundaries of \emph{invertible} phases which are non-trivial even in the absence of global symmetries. For example, the $\nu=1$ member of 1+1 d time-reversal invariant topological superconductors~\cite{FidkowskiKitaev_Z8_PhysRevB.83.075103} to which Kitaev's Majorana chain~\cite{Kitaev_majoranachain_2001} belongs. However, the unusual boundary Hilbert space of such phases demands a more careful investigation. Next, it is interesting to see how these results generalize to higher dimensions. While signatures of boundary SUSY (such as Bose-Fermi degeneracy) can be established in several higher dimensional examples, a general proof is currently lacking. It is also interesting to connect these results to the symmetry-extension framework of ref~\cite{WangWenWitten_PhysRevX.8.031048} as well as unwinding of SPT phases~\cite{APWangWei_Unwinding_PhysRevB.98.125108} in general dimensions. The one-dimensional case was recently studied in ref~\cite{AP_JW_Unwinding}. Finally, the emergence of \emph{space-time} SUSY on the boundary of topological phases has been investigated, first by Grover, Sheng and Vishwanath~\cite{GroverVishwanath_SUSY_280} and subsequently others~\cite{YaoLieaau1463,Yao_PhysRevLett.118.166802,Yao_PhysRevLett.119.107202}, when the boundary is tuned to criticality. It would be interesting to see in what way their results are related to ours. We leave these questions for future work.

\medskip 
\noindent \textbf{\emph{Acknowledgments}}: 
We are grateful to N.S. Prabhakar and R. Loganayagam for helpful comments and discussions. A.P. acknowledges funding from the Simon's foundation through the ICTS-Simons prize postdoctoral fellowship. J.W. is supported by the NSF Grant No. DMS-1607871 “Analysis, Geometry and Mathematical Physics” and by Center for Mathematical Sciences and Applications at Harvard University.

 \FloatBarrier

\bibliography{references}{}

\end{document}